\documentclass[onecolumn]{article}  
\usepackage{graphicx}	% Including figure files
\usepackage{amsmath}	% Advanced maths commands
\usepackage{amssymb}	% Extra maths symbols 
\usepackage{bm} 
\def\vlsr{v_{\rm LSR}}
\def\Msun{M_\odot}

\def\deg{^\circ}

\def\Xco{X_{\rm CO}}

\def\Tb{T_{\rm B}}

\def\htwo{H$_2$}
\def\Hcc{H cm$^{-3}$ }

\def\Msun{M_\odot}
\def\deg{^\circ}
\def\co{$^{12}$CO }
\def\coth{$^{13}$CO }
\def\coei{C$^{18}$O }
 
\def\kms{km s$^{-1}$}

\def\mH{m_{\rm H}}
\def\Tb{T_{\rm B}}  
\def\vlsr{v_{\rm lsr} }
\def\Xco{X_{\rm CO}}
\def\nhtwo{n_{\rm H_2}}
\def\htwocc{H$_2$ cm$^{-3}$}
\def\mum{$\mu$m }
\def\red{}
 
\title{Dark Supernova Remnants revealed by CO-line Bubbles in the W43 Molecular Complex along the 4-kpc Arm}
 
\author{Yoshiaki Sofue\thanks{E-mail: sofue@ioa.s.u-tokyo.ac.jp} \\ 
Institute of Astronomy, The University of Tokyo, Mitaka, Tokyo 181-0015, Japan }

\date{Accepted for Galaxies, 2021}  
 
\begin{document} 
\maketitle  
%%%%%%%%%%%%%%%%%%%%%%%%%%%%%%%%%%%%%%%%%%%%%%%%%%%%%%%%%         
\begin{abstract}   
Fine structure of the density distribution in giant molecular clouds (GMC) around W43 (G31+00+90 \kms at $\sim 5.5$ kpc) was analyzed using the 
FUGIN$^*$ CO-line survey at high-angular ($20''\sim 0.5$ pc) and velocity (1.3 \kms) resolutions ($^*$Four-receiver-system Unbiased Galactic Imaging survey with the Nobeyama 45-m telescope).
The GMCs show highly turbulent structures, and the eddies are found to exhibit spherical bubble morphology appearing in narrow ranges of velocity channels. 
The bubbles are dark in radio continuum emission, unlike usual supernova remnants (SNR) or HII regions, and in infrared dust emission, unlike molecular bubbles around young stellar objects.
The CO bubbles are interpreted as due to fully evolved buried SNRs in molecular clouds after rapid exhaustion of the released energy in dense molecular clouds.
The CO bubbles may be a direct evidence for exciting and maintaining the turbulence in GMCs by SN origin.
Search for CO bubbles as "dark SNRs" (dSNR) will have implication to estimate the supernova rate more accurately, and hence the star formation activity in the Milky Way. 
\end{abstract}    
 
 
%%%%%%%%%%%%%%%%%%%%%%%%%%%%%%%%%%%%%%%%%%%%%%%%%%%%%%%%

\section{Introduction}

Galactic supernova remnants (SNRs) are observed as extended objects bright in radio, X-ray, and/or optical emissions, often exhibiting shell structures expanding at high velocities 
\cite{chevalier1977,raymond1984,weiler+1988}.
About 300 Galactic SNRs are currently catalogued in the Milky Way 
\cite{green2009,green2019}. 
Their feedback to the ISM via interaction with the ambient molecular clouds has crucial implication to the interstellar physics such as the origin of interstellar turbulence, star formation, and cosmic ray acceleration \cite{wheeler+1980,cox1999,seta+2004,inoue+2012,sofue+2020snr}.
 
\red{Although most of radio SNRs are supposed to be catalogued for the transparency of the Galactic disc in radio, a larger number of SNRs is predicted from the estimated supernova rate of $0.02\pm 0.01$ SNe y$^{-1}$ \cite{tamman+1994,prantzos+1996,binns+2005,diel+2006,diel+2006b,{nath+2020}}, suggesting 200 to 2000 SNRs in the Galaxy for supposed life time of a SNR of $10^{4-5}$ y.}
Since remnants of core collapse SNe are expected to be located near to their birth places because of the short lifetime of high-mass progenitors, it is expected that their distribution is tightly correlated with that of HII regions.
However, only a weak concentration has been found in the spiral arms for the known SNRs \cite{li+1991}. 
In order to uncover missing SNRs, an extensive survey has been obtained by radio continuum observations, and a large number of candidate SNRs have been detected \cite{anderson+2017}.
However, their longitudinal distribution does not indicate a clear correlation with that of HII regions.

From these statistics we may expect that there exist a larger number of uncovered SNRs that are not detected in the current observations in radio or in other wave lengths.
A possible way of existence of such uncovered SNRs would be buried SNRs in molecular clouds (MC).
Supernovae (SNe) exploded in dense gas of MCs evolve in a quite different way from that exploded in the low-density inter cloud space.
They evolve rapidly in a short lifetime of $\sim 10^2$ y reaching a small radius of a few pc after peaky infrared flash \cite{shull1980,weiler+1988,lucas+2020,gupta+2020}.
Hence, their direct detection is difficult for the short time scale and compactness, so that no observational evidence has been obtained yet of the buried SNRs. 

In spite of the short flashing phase, the molecular cavity left after the buried SNR can survive for much longer time, which would be detectable as a cavity or a bubble of molecular gas.
We recently reported the discovery of an almost perfect round-shaped cavity with a clear-cut boundary in a medium sized molecular cloud at G35.75-0.25+28 in the \co ($J=1-0$) line emission \kms \cite{sofue2020dSNR}. 
The cavity is quiet in radio emission, unlike usual SNRs or HII regions.
It is also quiet in infrared emissions, unlike Spitzer bubbles associated with molecular shells around young stellar objects (YSO) \cite{chur+2006,deharveng+2010}. 
The peculiar property of the molecular cavity G35.75 was understood as due to a relic of a fully evolved supernova remnant (SNR) in the cloud, which may be the first evidence for the existence of a buried SNR in the Galactic disc.
We called the molecular cavity a "dark SNR" (dSNR).

In this paper, we extend the search for dSNRs in the form of molecular bubbles and/or cavities in the giant molecular clouds (GMC) surrounding the active star-forming region, W43 Main (G31+00+90 \kms), in the tangential direction of the 4-kpc molecular arm.
The molecular gas properties and star formation in the W43 region have recently been extensively studied using the FUGIN CO line data \cite{sofue+2019,kohno+2020}. 
We adopt a distance of 5.5 kpc according to the literature, which is close to the near-side kinematical distance of $5.56\pm 0.46$ kpc at $\vlsr=93$ \kms (center velocity of W43) for the most recent rotation curve of the Galaxy \cite{sofue2020rc}.

We make use of the FUGIN (Four-receiver Unbiased Galactic Imaging survey with the Nobeyama 45-m telescope) survey data in the $^{12}$CO, $^{13}$CO, and C$^{18}$O line emissions \cite{ume+2017}.
The data are available at the URL of FUGIN, http://nro-fugin.github.io.
The full beam width at half maximum of the 45-m telescope was $15''$ at the \co $(J=1-0)$ frequency and the velocity resolution was 1.2 \kms.
The effective beam size in the used 3D FITS cube is 20$''$, rms noise level $\sim 1$ K, and the pixel size  ($\Delta l, \Delta b, \Delta \vlsr) = (8''.5, 8''.5, 0.65$ \kms). 
 
%%%%%%%%%%%%%%%%%%%%%%%%%%%%%%%%%%%%%%%%%%%%%%
\section{CO Bubbles}

\subsection{Positions, Sizes and Distribution}

 In figure \ref{ch4} of the Appendix, we show channel maps of the \co-line brightness temperature in the $2\deg \times 2\deg$ region around W43.
Figure \ref{us20} shows the same, but extended emissions with scale sizes greater than $3' (\sim 5$ pc) are subtracted in order to enhance shells, arcs, and/or filaments.
All the channel maps exhibit numerous bubbly features (hereafter, bubbles), arcs and filaments.
Bubbles and arcs show up in the channel maps more pronounced than in the integrated intensity map of the same region \cite{kohno+2020}.
Each of the bubbles appears in a narrow range of velocity of a few \kms, and the diameter is typically $\sim 0\deg.1$ (10 pc), ranging from $\sim 2'$ to $\sim 15'$ (3 to 24 pc).
Arcs and filaments are generally fainter than complete bubbles, and are more extended.

\begin{figure}     
 	\begin{center} 
\includegraphics[width=16cm]{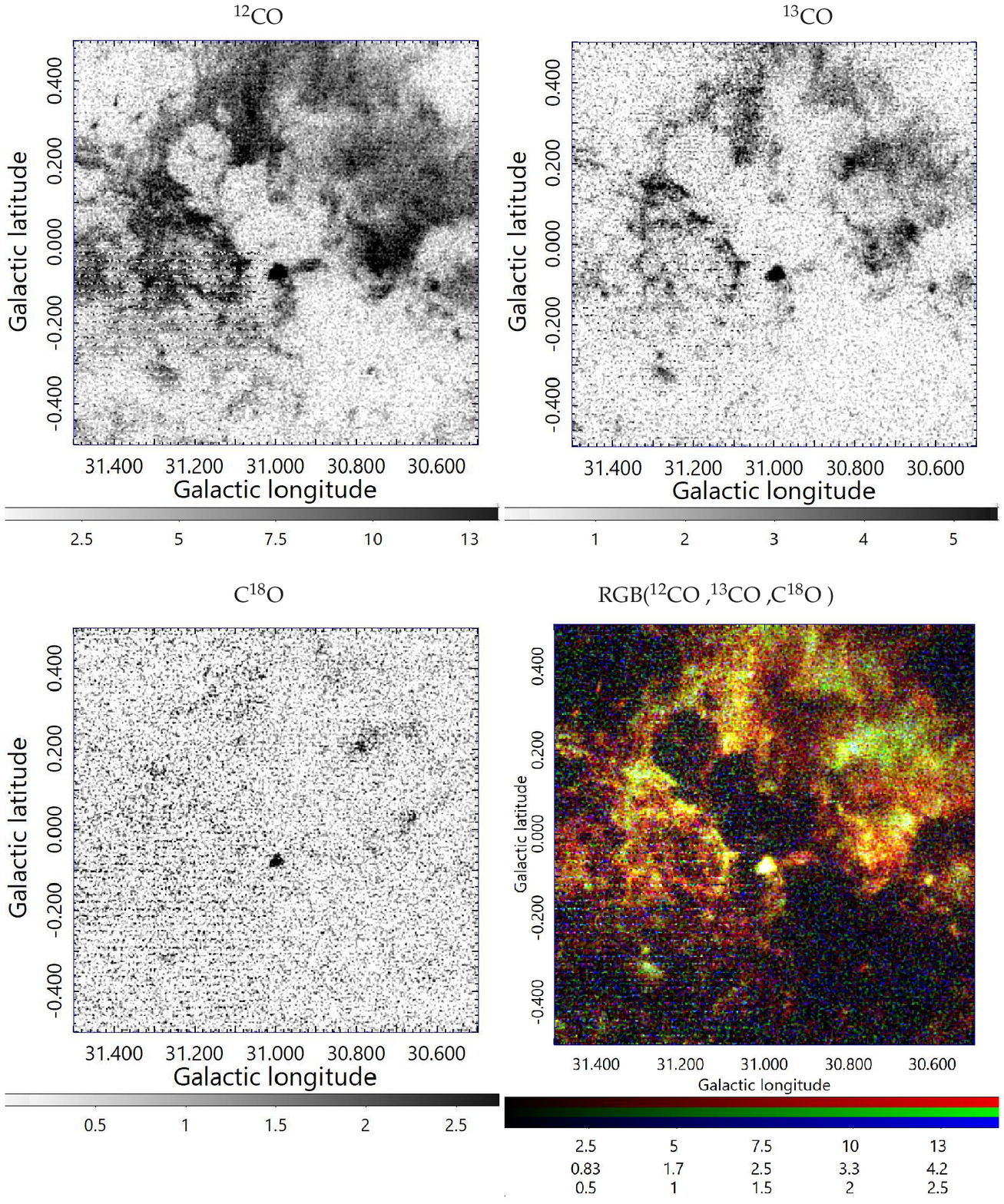}
\end{center}
 \caption{Brightness temperature maps of the G31+00 region at 81.675 \kms in the \co, \coth, and \coei $(J=1-0)$ lines, and a color-coded (RGB) map of \co (red, 0 - 15 K), \coth (green,0 - 5 K), and \coei (blue, 0 - 3 K).} 
\label{rgb}    
\end{figure}

Figure \ref{rgb} shows typical cases from a channel map, where CO bubbles at G31.2+0.2+81 at $\vlsr= 81.465$ \kms are shown by a composite color-coded map of \co brightness in red ($\Tb$ from 0 to 15 K), \coth in green (0 to 4 K), and \coei in blue (0 to 2 K).
Several almost empty cavities, each about $0\deg.1$ in diameter, are aligned from G30.8-0.2 to G31.2+0.2, apparently in touch with each other.
 The surrounding molecular gas makes shell structures with enhanced density.
Since there is no signature of excess in green intensity (\coth line), the molecular gas is mildly compressed at the bubble edges.  

 Using the original channel maps of a $2\deg\times 2\deg$ region around W43 taken from the FUGIN FITS cube data (http://nro-fugin.github.io), we identified many bubbles and arcs as shown in figure \ref{bub-plot}.
Their positions, approximate radii, center velocities, and approximate velocity widths are listed in table \ref{tab-bubbles}.

%%%%%%%%%%%%%%%%%%%%%%%%%%%%%%%%%%%%%%%%%%%%%%
\subsection{Quiet in radio and far infrared}

 In figure \ref{bub-plot} we plot positions and extents of  extended radio continuum sources at 5.8 GHz taken from the VLA survey (GLOSTAR) 
 \cite{medina+2019} with a similar angular resolution ($18''$) as the present CO maps. 
Except for W43 Main associated with comparable sized shells in CO and radio continuum, there appear no clear corresponding pairs. 

\begin{figure} 
 	\begin{center} 
\includegraphics[width=15cm]{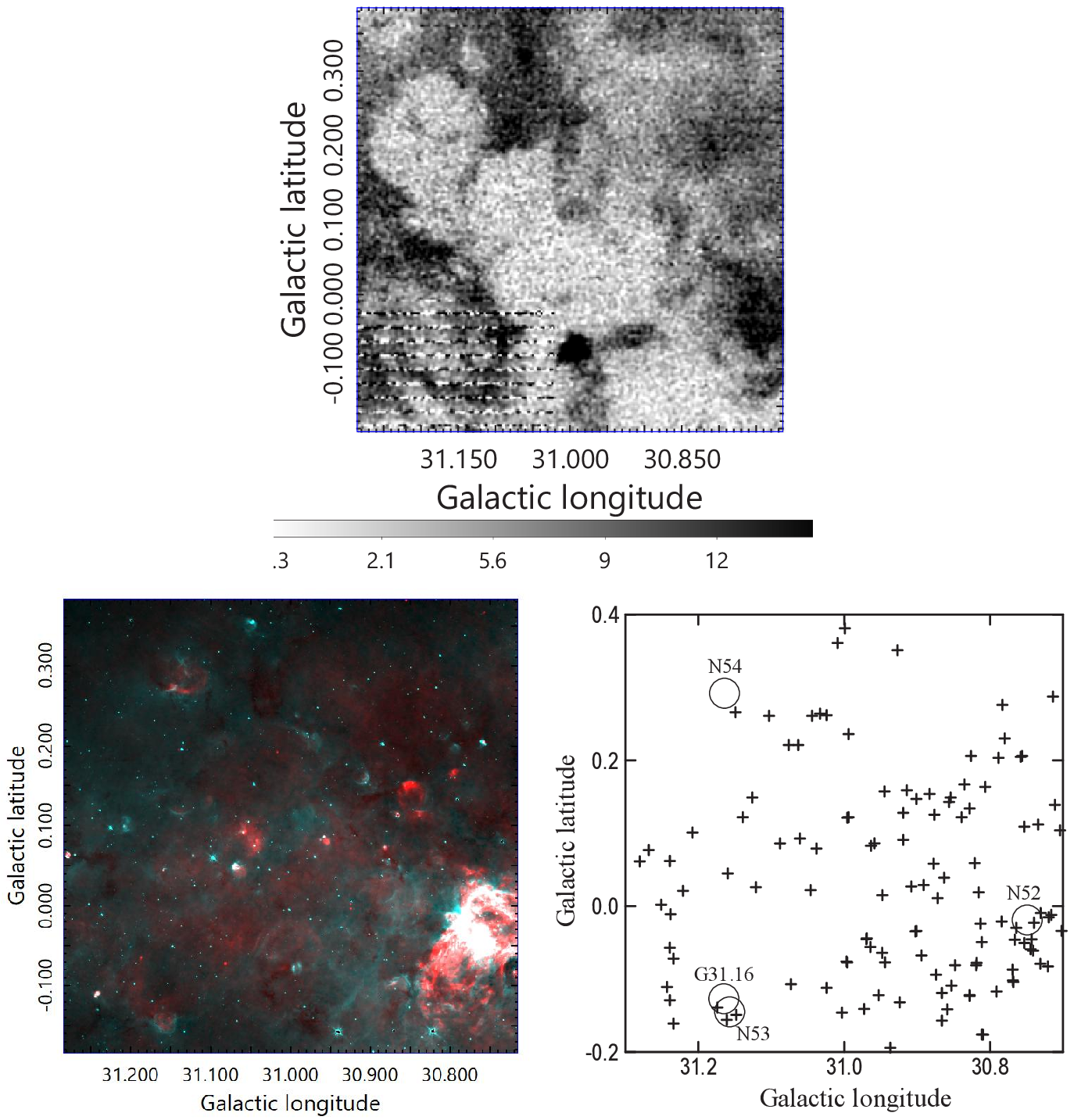}  
\end{center}  
\caption{
(Top) CO bubbles around G31+0.1 in \co $\Tb$ (K) at 81.465 \kms, 
(bottom left) 20 cm brightness in red (from 0 to 10 mJy/beam; MAGPIS) and 8\mum brightness  in blue/green} (from 50 to 500 mJy/str; GLIMPSE), and
(bottom right) YSO positions by crosses (SIMBAD: AGAL, 2MASS; http://simbad.u-strasbg.fr/simbad/) and Spitzer bubbles by circles with names.% \cite{deharveng+2010} (not size),    
\label{comparison}    
\end{figure}

In figure \ref{comparison} we enlarge the bubbles around G31.0+0.1, and compare with the radio continuum emission at 20 cm \cite{chur+2009,helfand+2006} (MAGPIS) and dust emission at 8 $\mu$m (GLIMPSE\footnote{http://www.spitzer.caltech.edu/glimpse360/aladin}).
We emphasize that the bubbles are not associated with radio continuum emission, unlike usual SNRs or HII regions.
Also, unlike molecular bubbles around young stellar objects (YSO), they are dark in thermal radio and infrared emissions.  

Panel (c) of the figure shows the positions of YSOs by crosses (SIMBAD) and Spitzer bubbles by circles (showing approximate extents)
\cite{deharveng+2010}, which are not coincident with the CO bubbles. 
These facts indicate that the CO bubbles are empty not only in the molecular gas, but also in warm dust, ionized thermal gas and non-thermal emitters (cosmic rays and magnetic fields).  
These CO bubbles are recognized only in the subsequent several channels (0.65 \kms increment), indicating that the velocity width of each bubble is several \kms. 

\subsection{Kinematics}

\begin{figure} 
 	\begin{center}  
\includegraphics[width=14cm]{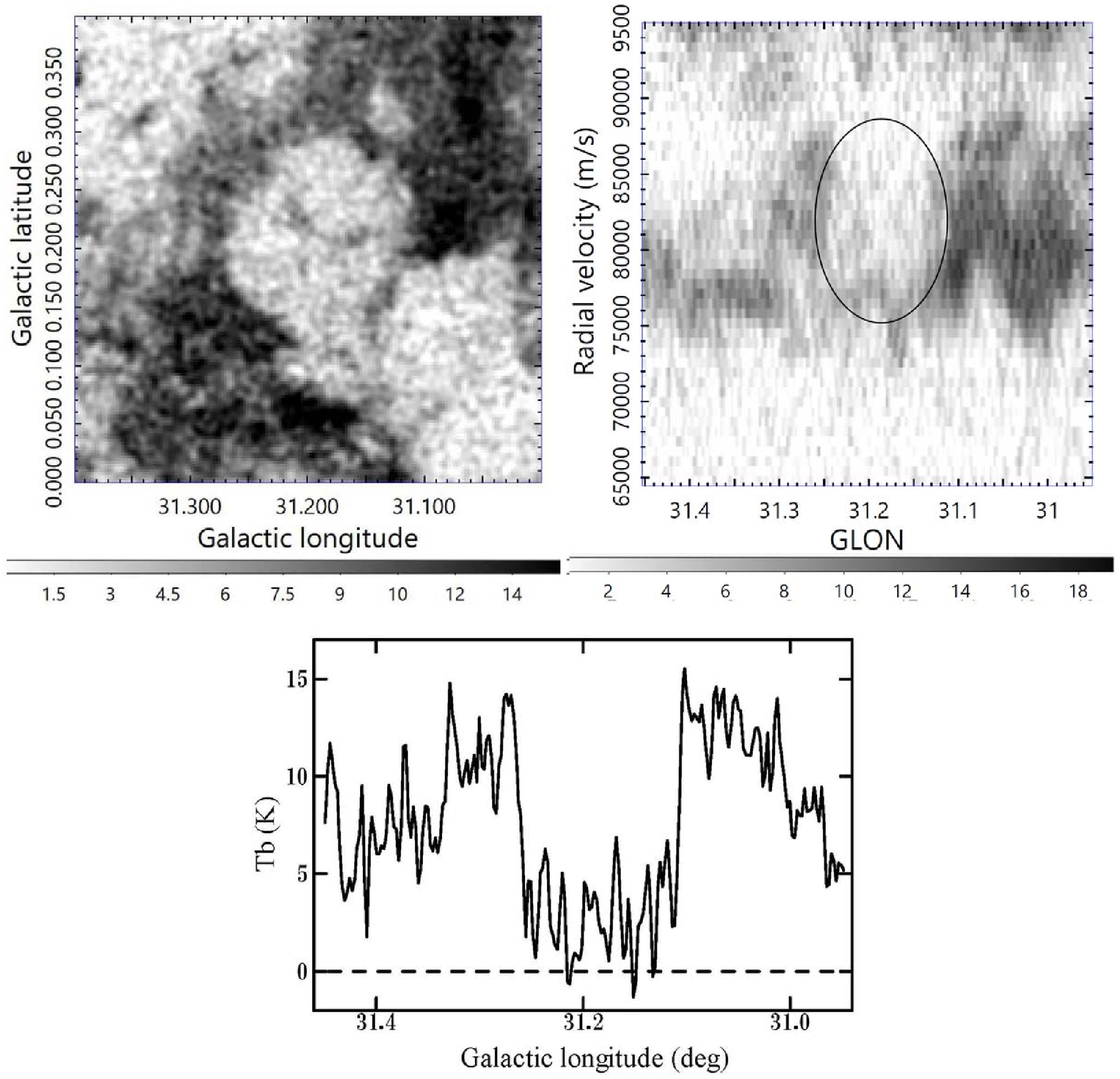} 
\end{center} 

\caption{(Top left) Molecular bubble S31.2+0.2 at 81.465 \kms.
(Top right) Longitude-velocity diagram around $b=0\deg.23$ across the bubble G31.2+0.2 (average of 4 channels in the latitude direction near the bubble center).
The ellipse represents bubble radius $0\deg.074$ (7.1 pc at 5.5 kpc distance) and half velocity width 6.7 \kms centered on $l=31\deg.19$, $\vlsr=81.9$ \kms.
(Bottom) $\Tb$ cross section at 81.0 \kms across the bubble center at position angle of $120\deg$.}
\label{s31.2+0.2_enlarge}    
\end{figure}

Figure \ref{s31.2+0.2_enlarge} shows a close up of the bubble at G31.2+0.2+81.675 \kms, and an averaged LV diagram across the center made from four subsequent LV diagrams.
The bubble is clearly visible as an elliptical ridge in the LV diagram as marked by an ellipse of radius of $\sim 0\deg.075$ and half velocity width of $\sim 7$ \kms.
Such an elliptical LV feature can be naturally understood as due to an expanding shell of radius 7.2 pc at velocity 7 \kms.
It is stressed that the thus estimated expanding velocity, which is ubiquitous in other bubbles, is greater than the velocity dispersion of a few \kms in the surrounding MC.
If the bubbles are dark SNRs, or the relics of buried SNRs, such increased velocity would be a direct evidence for the acceleration of interstellar turbulence by the feedback kinetic energy of an SN explosion.

In the third panel of figure \ref{s31.2+0.2_enlarge}, we present a cross section of $\Tb$ across the bubble center.
It reveals a clear-cut inner wall with the intensity maximum at the edge followed by an extended outskirt.
 This indicates that the bubble is a vacant cavity, 
and suggests that the interior gas, which had filled the cavity, is accumulated near the edge of the cavity, composing a shell structure observed as a CO bubble.
%already exhausted as being transformed to stellar mass during the star cluster formation, one of which was the SN progenitor. 
Alternatively or additionally, the inner gas may have escaped from the bubble through a smaller hole or a crack in the wall.
In fact, the bubble is not perfectly surrounded by the wall, but some parts are missing, being merged with the neighbouring bubbles.

%In figure \ref{s31.2+0.2_ch+lv} we show longitude-velocity (LV) diagrams across several bubbles around around G31.2+0.2. The LV diagrams across these bubbles also show similar elliptical features elongated in the velocity direction for $\sim 5-10$ \kms.

\subsection{Are CO bubbles ubiquitous?}

In order to examine a wider area for the CO bubbles, we show channel maps of the $2\deg \times 2\deg$ region around W43 in the Appendix, along with background-filtered images of each channel map to enhance bubbly and filamentary features.
Thereby, we detected many possible bubbles as indicated in the Appendix.

\begin{figure*} 
\begin{center}  
\includegraphics[height=17cm]{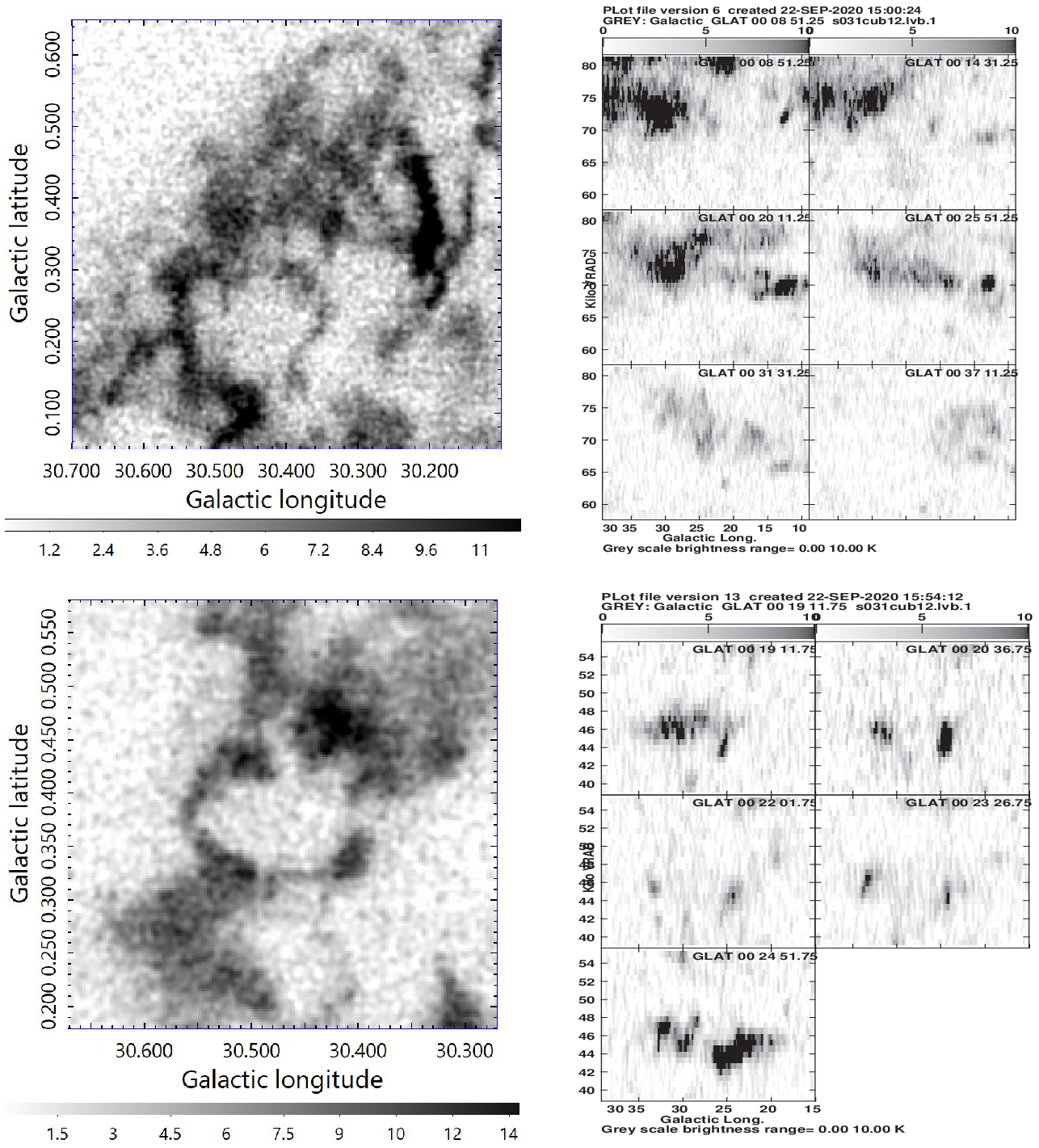}  
\end{center}
\caption{$\Tb$ maps and LV diagrams of CO bubbles at different velocities in different arms
: (Top) G30.4+0.4+70 \kms and (bottom) G30.45+0.36+46 \kms.}
\label{variousbubbles}    
\end{figure*}

Although the purpose of this paper is to search for CO bubbles in the molecular complex around W43 between 80 and 100 \kms, it may be worthwhile to look for similar objects in different regions and velocity ranges. 
Figure \ref{variousbubbles} shows examples of such bubbles found at different velocities, and hence in different arms, at G30.4+0.4+70 \kms and G30.45+0.36+46 \kms in the same sky area as for W43. 

G30.4+0.4+70 is located on the lower-velocity branch of the Scutum arm, and bubble diameters are about $0\deg.12$, corresponding to $D\sim  8.5$ or  19.6 pc for near or far distances of $4.2\pm 0.3$ and $9.6\pm 0.3$ kpc, respectively. 
The LV behavior is not straightforward due to superposition of multiple  bubbles, showing some vertical (velocity) elongations suggesting expansion at several \kms.

G30.45+0.36 +46 \kms is located on the higher-velocity branch of the Sagittarius arm. 
The radius of $0\deg.12\times 0\deg.1$ corresponds to 6.1 or 22.4 pc for near and far distances of $2.9\pm 0.3$ and $10.9\pm0.3$ kpc, respectively.
The LV diagram shows an elliptical feature, indicating an expanding motion at several \kms.
If we adopt the near distances, the sizes are comparable to those obtained for W43 bubbles.

The present search for CO bubbles was obtained only for the limited area on the sky around G31+00.
However, it is expected that similar CO bubbles would be found generally in many other molecular clouds, particularly in giant molecular clouds adjacent to star forming regions.
We may, therefore, reasonably argue that the CO bubbles of the same property as found here are rather ubiquitous in the Galactic disc, particularly in molecular complexes near active star forming regions in dense spiral arms.

% 	\begin{figure*}     
% 	\begin{center}   
%\includegraphics[width=16cm]{fig4.eps}  
%%\includegraphics[width=10cm]{fig4-s31.2+0.2_ch.eps}  
%\includegraphics[width=10cm]{fig4-s31.2+0.2_lv.eps}  
%\end{center}
%\caption{(Left) Channel maps of \co brightness temperature, and (right) longitude-velocity diagrams around G31.2+0.2. }    
%\label{s31.2+0.2_ch+lv}
%\end{figure*}     

\subsection{Mass and energy}

It is not easy to estimate the mass of a bubble, not only because it is vacant of the gas, but also because the outer boundary of the molecular cloud is not definite for the extended outskirt.
Instead, we may calculate the mass of supposed exhausted molecular gas that had filled the bubble in the past, assuming that the density was same as the ambient cloud density. 
The mean $\Tb$ of the surrounding gas cloud around the bubble G31.1+0.2+81 is about $\Tb \sim 10$ K, velocity width is $\delta v \sim 5$ \kms.
The mean molecular density can be estimated to be 
$\nhtwo \sim \Xco \Tb \delta v/r \sim 230$ cm$^{-3}$, where $r=7$ pc is the bubble radius and $\Xco\sim 2\times 10^{20}$ \htwocc (K \kms)$^{-1}$ is the conversion factor for extended molecular clouds \cite{sofue+2020xco}.
The total lost mass inside the bubble is then estimated as
$M \sim 4 \pi/3 \mu \nhtwo \mH r^3 \sim 2.3\times 10^4 \Msun$, where $\mu=2.8$ is the mean molecular weight and $\mH$ is the hydrogen mass.
If the mass had escaped from the bubble at a velocity of $v_{\rm esc}\sim 7$ \kms, the total kinetic energy of the thus lost mass is
$E_{\rm esc}\sim 1/2 M v_{\rm esc}^2 \sim 10^{49}$ erg, safely supplied by the input energy by an SN explosion.
 
%%%%%%%%%%%%%%%%%%%%%%%%%%%%%%%%%%%%%%%%%%%%%%
\section{Discussion}

\subsection{More Bubbly GMCs than Filaments}

We showed that the giant molecular clouds (GMCs) composing the W43 molecular complex are filled with CO bubbles.
Such bubbly structures are particularly evident in the channel maps after subtracting the extended emission (figure \ref{us20}).
 Thus, in so far as the W43 complex is concerned, the GMCs are generally bubbly rather than exhibiting filament structures.
This makes contrast to the filamentary interstellar turbulence in the local Orion clouds  
\cite{bally+1987} or to that expected from simulations 
\cite{federrath+2010}. 
Hence, as will be concluded in the next subsection, the bubbly behavior may be due to a more efficient feedback by the current SF activities in W43 associated with a larger number of SNe compared to that in the local ISM in the solar vicinity.   

\subsection{ Origin of CO Bubbles} 
 
 As to the origin of the CO bubbles, we may consider several possible mechanisms.\\
(i) The first idea is that they are fully evolved relics of buried SNRs, simply argued from the required total energy. Thereby, the shell structure is maintained by its own expanding motion, as described later.\\
(ii) The second idea is that they are evolved Spitzer bubbles.
This idea  encounters the difficulties, as raised in the previous section, that the bubbles are quiet in thermal radio and infrared emissions.\\
(iii) Stellar winds from young stars may also be excluded, because there is no signature of star formation inside the bubbles, as for the reasons against (ii), except for the core area of the W43 Main.\\
(iv) Outflows from old population stars such as planetary nebulae would be another possibility \cite{balick+2002}. 
The responsible mass-loss stars are distributed in the population II disc at a number density approximately equal to that of AGB stars 
$ 
n_{\rm PN}
\sim t_{\rm AGB}/t_* (M_{\rm disc}/M_*) /(\pi r_{\rm disc}^2 z_{\rm disc})
\sim 10^{-6} {\rm pc}^{-3},
$
where $t_{\rm AGB}\sim 10^3$ y, $t_*\sim 10^{10}$y, $M_{\rm disc}\sim 10^{11}\Msun$ is the Galactic disc mass, $M_*\sim 1\Msun$, $r_{\rm disc}\sim 5$ kpc and $z_{\rm disc}\sim 200$ pc is the disc radius and full thickness, respectively. 
We thus expect only one such star within $\sim 100$ pc volume around W43. 
Moreover, the supplied energy would be too small, 
$E_{\rm PN}\sim (1/2)v^2 \dot{M} t_{\rm AGB} \sim 10^{45}$ ergs from the wind, where $\dot{M} \sim 10^{-8}\Msun {\rm y}^{-1}$ is the mass-loss rate and $v\sim 2500$ \kms is outflow velocity.\\
(v) Thermal instability produces a cavity, if the heating rate by cosmic rays per molecule is constant and the cooling rate is proportional to the square of gas density \cite{sofuesabano1980}.
A lower-density perturbation results in a growing cavity. 
However, it cannot explain the observed expanding velocity of the shell at several \kms, because the perturbation grows at the sound speed of molecular gas, $\sim 1$ \kms. \\
(vi) Magnetic filaments will produce perpendicular molecular filaments \cite{tahani+2019,gomez+2018}. 
However, in order to make CO bubbles, the magnetic fields must be radial about the bubble centre.\\
(vii) Finally, one may attribute the bubbles to interstellar turbulence. 
However, this argument does not answer the question about the origin of the CO bubbles. 
In fact, ideas (i) to (vi) are almost equivalent to that about the origin of turbulence in molecular clouds. 

\subsection{Dark SNRs}

 From the above consideration, we here conclude  that idea (i) is most plausible as the origin of the observed CO bubbles.
We here try to explain the CO bubbles by well evolved and radio quiet SNRs, which exploded inside molecular clouds and had evolved as buried SNRs.
We assume that the responsible energy sources are mostly core-collapsed (type II) SNe, because most of the catalogued SNRs, mainly from radio observations by their shell structures, are of type II SN origin.
Type Ia SNe would make SNRs of filled center morphology, while rarely produce shell structures.
Also, kinetic energy released by this type is not sufficient to explain the expanding kinetic energy of the CO bubbles. 

 Massive stars produce cavities in the ambient gas by the stellar winds for some My.
The wind-driven shells evolves into shocked SNRs soon after SN explosions \cite{gupta+2020,lucas+2020}.
By scaling the current SNR models %simulations \cite{gupta+2020}
for ambient density of $\sim 1$ \Hcc to a case of $\sim 10^3$ \Hcc in a MC, both the radius and velocity can be scaled down by a factor of $100^{-2/5}=0.16$ for the same time scale unit.
When the shock wave reaches the wind's boundary, the molecular gas is compressed and evolves as a buried SNR.
Here, we consider a case that massive stars are distributed over the GMC, and they end their lives as individual SNe. 

 If high-mass stars compose a dense cluster ending by multiple SNe, they will disrupt the ambient clouds \cite{gupta+2020}.
In this case, the SNe may not leave such a bubbly GMC as observed around W43, but will blow off the surrounding MCs, from which they formed, leaving a naked stellar cluster.  

The expansion velocity $v$ and radius $r$ of a spherical adiabatic shock wave in a uniform-density gas are related to the input kinetic energy $E_0$ and gas density $\rho_0$ as 
$E_0 \sim (1/2)(4 \pi/3) r^3 \rho_0 v^2$, 
where $\rho_0=\mu n_{\rm H_2, 0}\mH$ is the ambient gas density. 
Most of the released energy by core-collapse SN explosion, 
$\sim 10^{51}$ erg, in a dense gas cloud is exhausted by the initial infrared flash within $\sim 10^2$ years \cite{shull1980,weiler+1988,lucas+2020}.
After the initial radiation phase, the kinetic energy given to the gas expansion may be assumed to be on the order of
$E_0\sim 10^{50}$ erg, an order of magnitude smaller than the total released energy.

%The relations between the expansion velocity $v$ radius $r$ and elapsed time from explosion $t$ of the shock wave front are plotted in figure \ref{sedov}, for three different energy-vs-density parameter $ED$.
We here introduce a parameter, $ED$, defined by
$
    ED={\rm log}(E_0/n_{\rm H}),
$
where $E_0$ is the input energy by the explosion in ergs, and $n_{\rm H}$ is the number density of hydrogen atoms in cm$^{-3}$.
The hydrogen number density in a molecular cloud is related to the \htwo density through $n_{\rm H}=\mu \nhtwo$ with $\mu=2.8$.
%For example, for $E_0=10^{51}$ erg and $n_{\rm H}=1$ H cm$^{-3}$ as for a typical SNR in the inter-cloud ISM, we have $ED=51$.For $E_0=10^{50}$ erg and $\nhtwo=10^3$ \htwo cm$^{-3}$ ($=2.8 \times 10^3$ H cm$^{-3}$), $ED=46.5$. 
The observed radius and velocity for the CO bubble G31.2+0.2 of $\sim 7$ pc and $\sim 7$ \kms is realized, when $ED=46.3$, and the age is determined to be $t\sim 0.4$ My. 
If we adopt $E_0=10^{50}$ erg, the density is required to be $n_{\rm H}=5\times 10^3$ H cm$^{-3}$, or  $\nhtwo \sim 2\times 10^3$ \htwocc.
 If the cooling is significant so that the input energy is equivalently decreased to $E_0=10^{49}$ erg, the density may be an order of magnitude lower, consistent with the measured density of $\sim 230$ \htwocc in the GMC around W43. 

% 	\begin{figure}     
% 	\begin{center}
%\includegraphics[width=14cm]{fig6.eps} 
%\end{center}
%\caption{Sedov solutions for energy-density parameter, $ED={\rm log} \ E_0/n_{\rm H}=46,47,$ and 48 from bottom to top curves.  Upper curves represent larger energy and lower density, and lower curves vice versa.A CO bubble with $v\sim 7$ \kms and $r\sim 7$ pc (straight lines) lies on curves with $ED\sim 46.3$.
%}
%\label{sedov}    
%\end{figure}       
   
\subsection{Evolution}

The presently identified molecular bubbles exhibit close resemblance to that reported in our earlier paper on G34.75-0.2 \cite{sofue2020dSNR}. 
We here try to explain the molecular bubbles as due to  dark SNRs, which had evolved in the molecular clouds as buried SNR in W43 molecular clouds, ceased their expansion, and faded out of the thermal and high-energy radiation phase.

The evolutionary time scale in the radiation phase of buried SNRs is two orders of magnitude shorter than the usual SNRs exploded in inter-cloud low-density regions because of the extremely higher ambient density, so that the luminous phase ends in $\sim 10-100$ y \cite{shull1980,weiler+1988,lucas+2020,gupta+2020}. 
For the short lifetime, they have little chance to be observed and catalogued as radio or optical SNRs, but can be recognized by molecular bubbles as dark SNRs in their latest phases.
 
Figures \ref{evolutionTop} illustrates the evolutionary scenario along a flow line of the Galactic rotation.
It schematically explains the spiral arm structures of molecular clouds, HII regions and of SNRs, according to the evolutionary scenario  under the galactic shock wave theory.
We may summarize the evolution from core collapse SNe to dSNR as follows.
%\ref{roberts1969}.
 
\subsubsection{Galactic shock wave, cloud collision, and star formation ($t\sim -1$ My)}

Diffuse ISM as well as molecular clouds are strongly compressed by the galactic shock wave along the 4-kpc molecular arm \cite{sofue+2019}. 
Due to both the galactic shock and orbital condensation in the bar-end, cloud collisions are strongly enhanced \cite{kohno+2020}.
Accordingly, intense star formation is activated at cloud interaction fronts, and OB stars are formed and HII regions are produced, emitting thermal radio and far infrared dust emissions.
A significant fraction of the formed OB stars and clusters develop inside the giant molecular cloud.
Frequent cloud collisions cause not only star formation, but also growth of molecular clouds by merger, resulting in formation of larger scale molecular complex. 

\subsubsection{SN explosion ($t=0$ y), buried, and cool SNR ($\sim 10^{1-2}$ y)}

The OB stars explode as supernovae, and their significant fraction are still embedded inside the molecular complex along the molecular arm.
Most of the released energy of SNe is exhausted by radiation of neutrinos, $\gamma$ rays, and hard-X rays.  
The ejecta of SNe and snow-plowed gas in the molecular cloud form expanding buried SNRs, which evolve rapidly due to strong cooling by the dust and thermal emissions in infrared and mm waves. 
The buried SNRs end their shining phase in a life time as short as $\sim 10^2$ y, leaving cool SNRs.

\subsubsection{Dark SNR as molecular bubbles ($\sim 10^3 - 10^5$ y)}

The evolved buried SNRs remain as dark SNRs, which expand as an almost adiabatic shock wave in the dense molecular gas.
They are observed as the molecular cavities and bubbles in their expanded phase, as reported in this paper.

 	\begin{figure}     
 	\begin{center} 
\includegraphics[width=11cm]{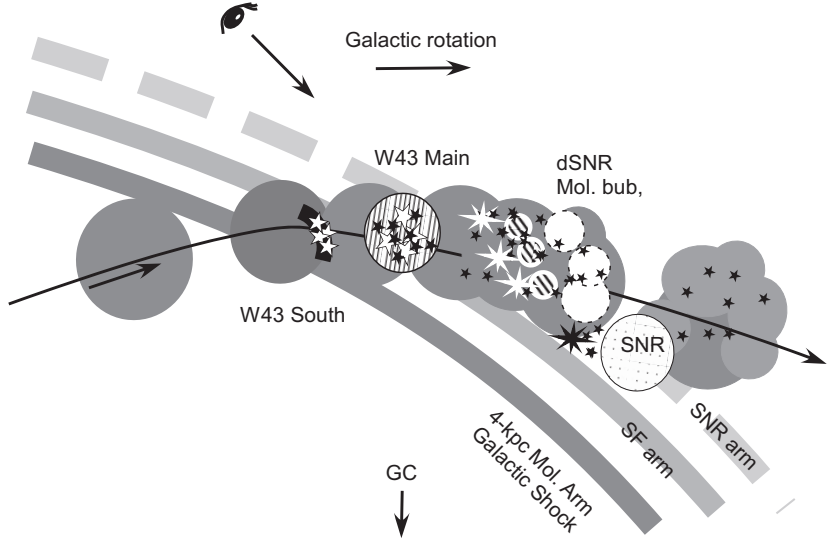} 
\end{center}
\caption{Face-on view of the 4-kpc arm from the north Galactic pole, illustrating the evolution of a molecular cloud and dSNRs along the Galactic flow line by rotation through a galactic shock. 
}
\label{evolutionTop}    
\end{figure}     

\subsection{Bubble properties}

 As readily shown in figures \ref{rgb}, \ref{ch4} and \ref{us20}, the analyzed region is full of molecular bubbles.  
The bubbles have a typical radius of $r\sim7$ pc, spreading from 5 to 15 pc.
Some arc shaped filaments with larger radii or length of 15 pc are also found in the maps, which are supposed to be segments and/or remnants of CO bubbles.
Besides those counted here, there appear a larger number of fainter shells and arcs that cannot be measured on the maps.
So, the here listed bubbles may be a minimal set of dark SNRs in the analyzed region. 
%The positions and approximate shapes of the identified shells and bubbles are superposed in the channel maps in figure \ref{ch4}.  
%Frequency distribution of the measured radii is plotted in figure \ref{bubbles}.
\begin{figure}     
 	\begin{center} 
\includegraphics[width=8cm]{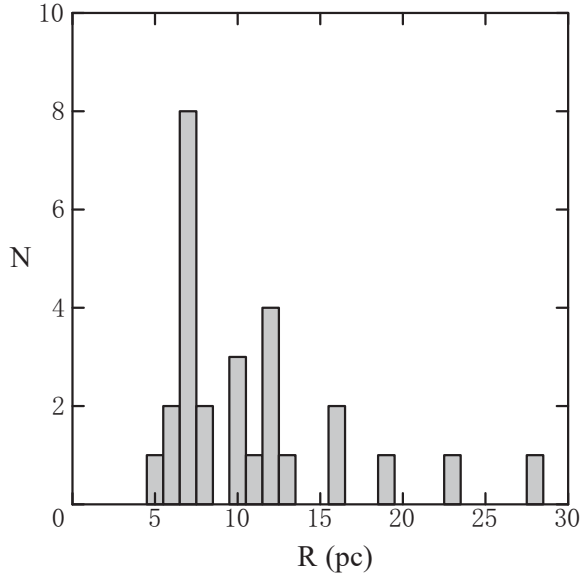}
\end{center}
\caption{Frequency of bubble radii for assumed distance of 5.5 kpc to W43.
}
\label{bubbles}    
\end{figure}      
%As seen from the comparison of the marked loops in figure \ref{ch4} with the bubbly features in the BGF maps, there appear to exist a more number of fainter, unidentified loops and arcs. 

It may be stressed that the here found bubbles exhibit round shapes. If they are rings or loops, such generally round shapes are not expected, but they must exhibit elongated ellipses on the sky because of the higher probability of viewing a ring obliquely.
It is, therefore, natural to consider that they are spherical bubbles. 

It is also emphasized that such molecular bubbles were recognized for the first time thanks to the high velocity and angular resolutions of the FUGIN survey.
In fact, they are hardly seen in the integrated intensity maps \cite{kohno+2020} or in lower resolution CO surveys \cite{dame+2001}. 
This is because each bubble appears in a narrow velocity range within a few \kms.

\subsection{SN rate in the Spiral Arm}
 
\red{The rate of SNe in the Galaxy has been estimated to be on the order of $2\pm 1$ per 100 y by various observations (see table 1 of \cite{diel+2006}).
Particularly, the rate of core collapse SNe supposed to be responsible for shell type SNRs has been rather accurately determined from the $\gamma$-ray spectroscopy \cite{prantzos+1996,diel+2006,diel+2006b}, yielding $1.9\pm 1.1$ per 100 y.
This predicts $\sim 200$ shell type SNRs in the Galaxy for an assumed life time of a shell of $\sim 10^4$ y, or $\sim 2000$ for $\sim 10^5$ y, strongly dependent on the adopted life time of a shell.
Estimation of the exact life time of a shell is difficult from observations of SNRs expanding into the turbulent ISM with significant deformation. 
Furthermore, the Galactic plane is observed to be full of unidentified radio filaments
\cite{stil+2006,chur+2006,helfand+2006},
suggesting the presence of a large number of debris of un-catalogued old SNRs.}

\red{We may thus argue that the density of existing shell type SNRs in the Galaxy is on the order of or greater than $\sim 200-2000$.
Furthermore, from the $^{26}$Al $\gamma$-ray spectroscopy and intensity distribution, the SNe have been shown to be concentrated around the Galactic Centre within $|l|\sim \le 30\deg$ \cite{diel+2006}. 
This means that we may expect a higher density of shell-type SNRs in the here studied region than $N\sim 200-2000/60\deg$ or 3 to 30 per degree of longitude.
Furthermore, if the SNRs are spatially correlated with the SF arms, they must be more concentrated in the tangential direction of the spiral arms.
Thus, we may expect a much higher, or the highest 
longitudinal SNR density in the tangential direction of the Scutum arm (4-kpc molecular ring) at $l\sim 30\deg$ nesting W43, the most active SF site in the first quadrant of the Galaxy.
}

\begin{figure} 
\begin{center}   
\includegraphics[width=16cm]{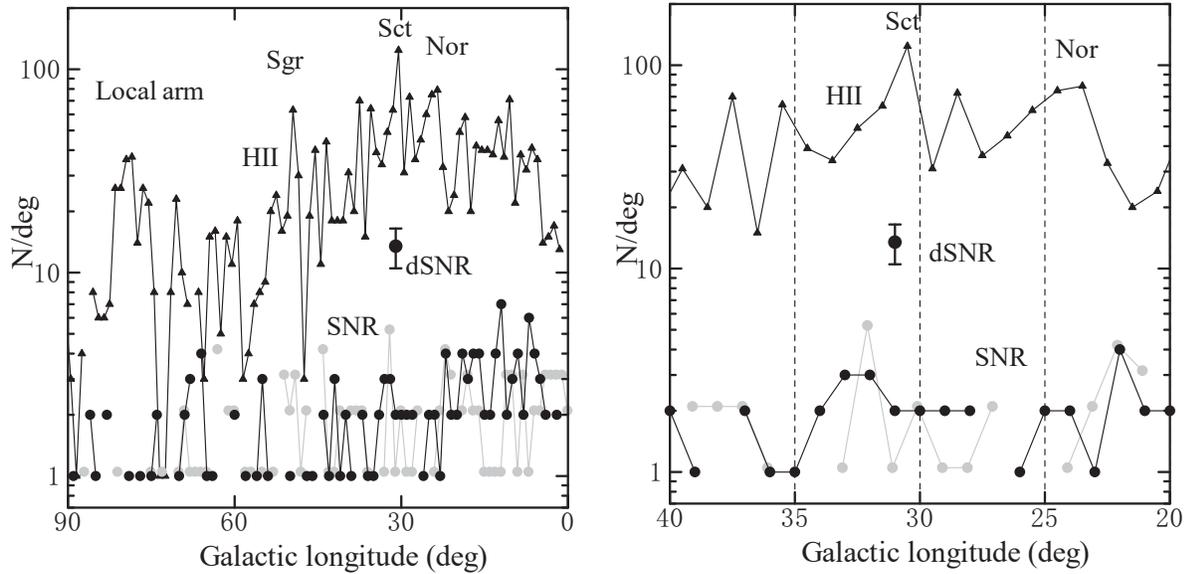} 
\end{center}
\caption{[Left] Longitudinal number density per one degree $N$ of Green's catalog SNRs (black and grey circles in the 1st and 4th Galactic quadrants, respectively), %{\cite{green2009}},
HII regions (triangles),
%\cite{hou+2014}}, 
and CO bubbles (dSNR) (big circle with error bars).
[right] Same, but enlarged around the Scutum Arm.}
\label{bubgreensta}    
\end{figure}

If the bubbles are relic of buried SNRs, the radii and expanding velocities suggest that their ages are on the order of $\sim 0.4$ My.
This is an upper limit to the age, and the real dSNRs would have evolved a bit rapider due to cooling effects.
So, we here assume that their ages are on the order of $10^5$ y.
 As counted in the Appendix, the number of CO bubbles in the longitude range from $l=30\deg$ to $32\deg$ is $N\pm \sqrt{N}=27\pm 5$ per 2 degrees in longitude, or $13.5\pm 3$ per degree.
On the other hand, the catalogued SNRs yields $N \sim 3$ per degree in the same direction. 
In figure \ref{bubgreensta} we plot the thus estimated counts in comparison with those of the catalogued SNRs \cite{green2009,green2019} as well as with the number of HII regions per degree \cite{hou+2014}.  

Although the longitudinal number density of the catalogued SNRs shows enhancement in the tangential directions of the spiral arms, it is significantly weaker than that of the HII regions. 
It may be also mentioned that the density peak of SNRs is about one degree shifted toward outer longitude side from the peak of HII regions. 
This means that the SNR arm is located outside the star-forming arm by about 100 pc, which  is consistent with the evolutionary scenario of SNRs in the spiral arm (figure \ref{evolutionTop}). 
 
We finally mention that an advantage of using CO bubbles is that the bubbles simultaneously yields radial velocities, and hence, kinematic distances of the dSNRs. 
Statistical analyses using a larger number of dSNRs with known distances would have implication to estimate the supernova rate in the spiral arms more accurately. 

%%%%%%%%%%%%%%%%%%%%%%%%%%%%%%%%%%%%%%%%%%%%%%
\section{Summary}
Numerous round-shaped bubbles and cavities of CO-line emission with radii of $\sim 5-10$ pc were found in the molecular complex around W43 (G31+00+90 \kms) in the tangential direction of the 4-kpc star-forming arm. 

The bubbles are quiet in radio continuum emission, unlike usual supernova remnants (SNR) or HII regions, and are dark in infrared dust emission, unlike molecular bubbles around YSOs.
The CO bubbles are interpreted as due to dSNR, or fully evolved SNRs buried in dense molecular clouds after rapid exhaustion of released energies by SNe.
Increased velocity width in the bubbles as seen in the LV diagrams compared with that in the ambient molecular gas may be a direct evidence for the acceleration of interstellar turbulence by SN explosions.

From the number count of the "dark" SNRs in W43 complex, we argue that the supernova rate currently estimated from the catalogued SNRs has been significantly under-estimated.
Such correction of the SN rate in the Galactic disc would affect the star formation history in the Milky Way.  
We proposed to use the CO bubbles to search for a more number of dSNRs.
Taking advantage of simultaneously obtained radial velocities, and hence kinematic distances to the dSNRs, we will be able to perform a more accurate statistical analyses of the correlation between SNR and HII regions in the Galaxy.\\
    
%%%%%%%%%%%%%%%%%%%%%%%%%%%%%%%%%%%%%%%%%% 
\noindent{\bf Acknowledgments}: The author is indebted to the FUGIN team for the archival data base of the CO line survey of the Galactic plane using the Nobeyama 45-m telescope.
The data analysis was partially carried out at the Astronomy Data Center of the National Astronomical Observatory of Japan.  The data underlying this article are available in the URL http://nro-fugin.github.io. \\

%%%%%%%%%%%%%%%%%%%%%%%%%%%%%%%%%%%%%%%%%%

\noindent{\bf Conflicts of interest}: The author declares no conflict of interest.  
%%%%%%%%%%%%%%%%%%%%%%%%%%%%%%%%%%%%%%%%%%
%\reftitle{References}

% Please provide either the correct journal abbreviation (e.g. according to the gList of Title Word Abbreviationsh http://www.issn.org/services/online-services/access-to-the-ltwa/) or the full name of the journal.

%\appendixtitles{yes} 
\begin{appendix}
\section{Channel maps and background-filtered images}  

\begin{figure}     
 	\begin{center} 
\includegraphics[width=15cm]{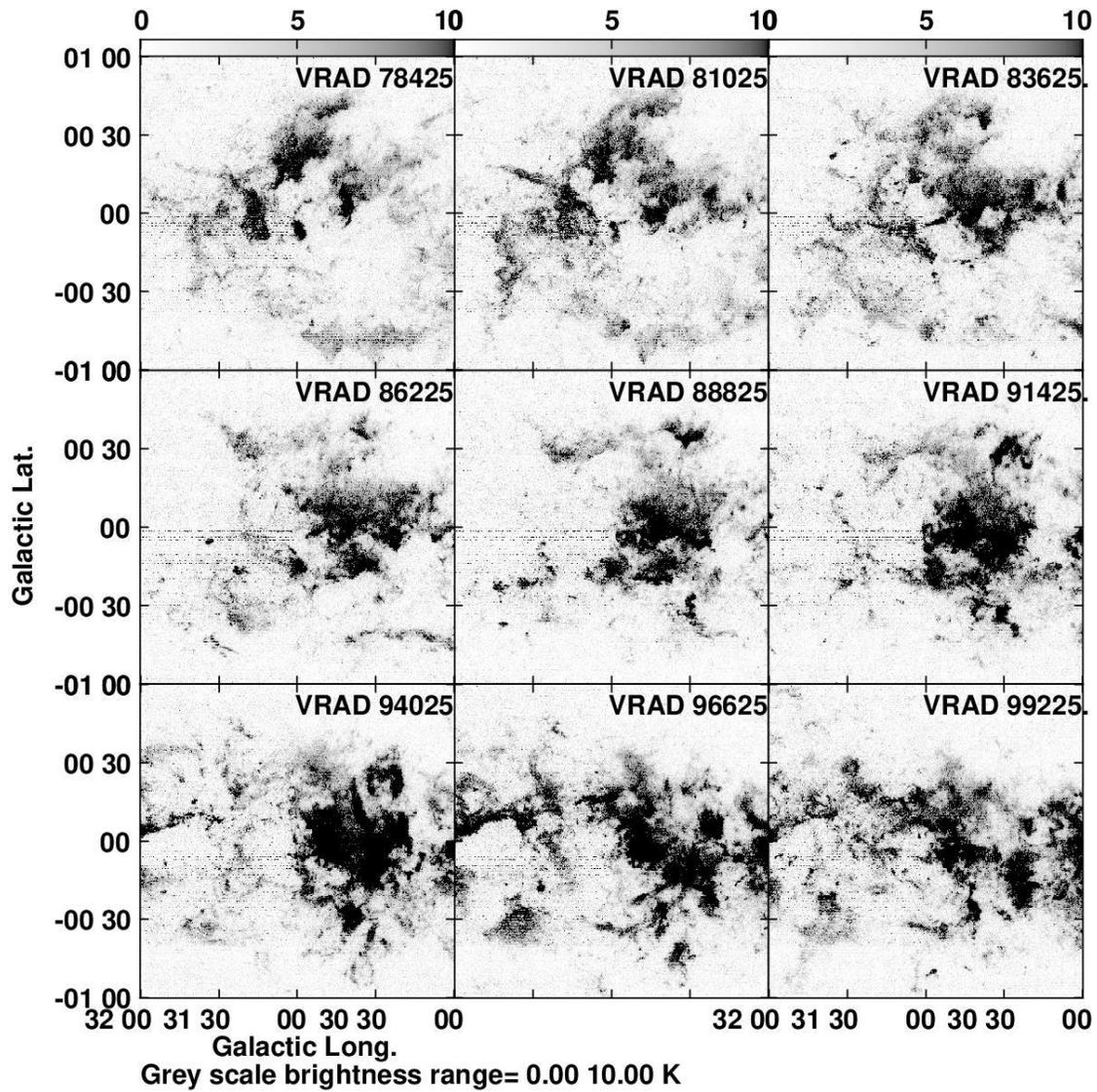}
\end{center}
\caption{Channel maps of \co brightness temperature of the W43 complex from 78 to 100  \kms every 2.6 \kms (every 4 original channels). 
 Radial velocities (VRAD) of the channels are indicated in m s$^{-1}$ in the individual panels. } 
\label{ch4}    
\end{figure}

\begin{figure}     
 	\begin{center}
\includegraphics[width=15cm]{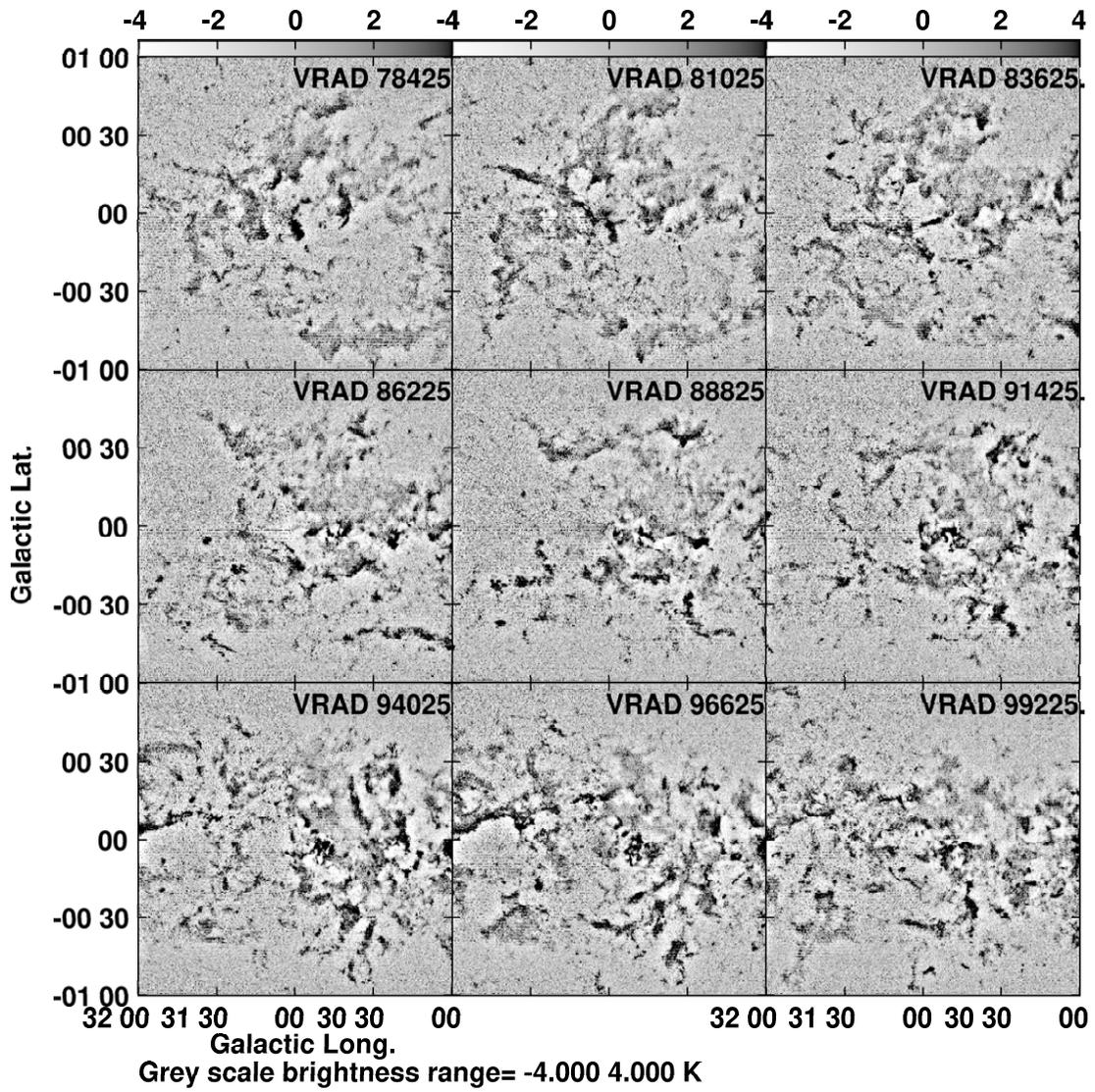}
\end{center}
\caption{ Same as Fig.\ref{ch4}, but extended structures have been subtracted in order to enhance shells and arcs.} 
\label{us20}    
\end{figure}

In figure \ref{ch4} we present channel maps of the \co \ brightness temperature, $\Tb$, in a $2\deg\times 2\deg$ region around W43 Main centered on G31+00+90 \kms every 4 original channels, or every $=4\times 0.65=2.6$ \kms velocity interval.
W43 Main is located at $l=30\deg.8,\ b=0\deg$ embedded in the giant molecular cloud at $\vlsr\sim 93$ \kms.
The used FITS cube data are available on the FUGIN web pages at  http://nro-fugin.github.io.
 Figure \ref{us20} shows the same, but extended structures with scale sizes greater than $\sim 3'$ have been subtracted in order to enhance smaller scale clouds and filaments. 
The figures exhibit numerous bubbles, arcs and filaments.

 Using the original channel maps between $\vlsr=80$ and 100 \kms at velocity increment of 0.65 \kms, we have traced CO bubbles by eye estimate.
Thereby, each bubble was identified as a loop of $\Tb$ ridge that can be traced over a couple of neighboring channels.
Their positions and radii are shown in figure \ref{bubbles}, and are listed in table \ref{tab-bubbles} along with radial velocities, $\vlsr$.
Half velocity widths of the bubbles, $\delta \vlsr ^{1/2}$, defined as half the velocity range in which a bubble can be traced on the neighboring channel maps, were also measured and listed in the table. 
It is shown that the expanding velocity of a bubble measured on the LV diagram is about three times the here listed half velocity width. 
We also list typical brightness temperature of each bubble edge as read on the relieved channel maps. 

 	\begin{figure}     
 	\begin{center}
\includegraphics[width=8cm]{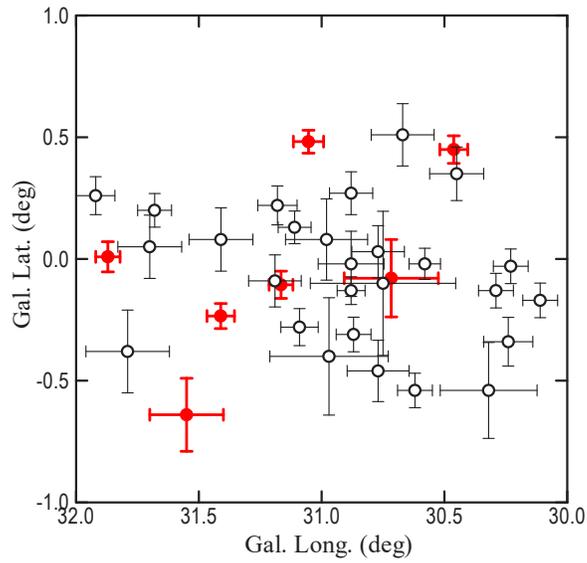}
\end{center}
\caption{ Open circles with thin bars indicate positions of CO bubbles traced in the channel maps between 80 and 100 \kms shown in figures \ref{ch4} and \ref{us20}, and approximate extent by the bar lengths. Thick red circles and bars show positions and extents of extended radio continuum sources at 5.8 GHz from the VLA observations \cite{medina+2019}. } 
\label{figA3}    
\label{bub-plot}
\end{figure}

\begin{table*}[htbp] 
\begin{center}
\caption{Parameters of CO bubbles} 
\begin{tabular}{ccccccccc} 
\hline
$l$&$b$& $r$& $r$&$\vlsr$&$\delta \vlsr ^{1/2}$ &$\Tb$\\%&s\\
(deg)&(deg)&(deg)&(pc)&(\kms) &(\kms) &(K)\\%&(\%)\\
\hline
30.11&   -0.17&   0.071&     6.8& 101.2& 3& 6\\%&30 \\
30.23&   -0.03&   0.071&     6.8 &82.3&3&4\\%&30\\ 
30.24&   -0.34&   0.100&     9.6 &86.6&2&7 \\%&30\\ 
30.29&   -0.13&   0.071&     6.8&94.7&2&9 \\%&40\\
30.32&   -0.54&   0.197&    18.9&88.8&3&9 \\%&60\\  
\\
30.45&    0.35&   0.110&    10.6&94.6&2&12 \\%&60\\ 
30.58&   -0.02&   0.064&     6.1& 83.0 &3& 6 \\%&100 \\
30.62&   -0.54&   0.071&     6.8&91.4&2&7 \\%&70\\ 
30.67&    0.51&   0.128&    12.3&83.6&2&4 \\%&50\\ 
30.75&   -0.10&   0.296&    28.4&84.9&4&6 \\%&50\\ 
\\
30.77&    0.03&   0.107&    10.3&79.2&2&6 \\%&50\\
30.77&   -0.46&   0.126&    12.1&96.6&3&5 \\%&40\\ 
30.87&   -0.31&   0.071&     6.8&99.2&3 & 4 \\%&30\\
30.88&   -0.02&   0.134&    12.8 &79.7&4&9 \\%&40\\
30.88&   -0.13&   0.057&     5.5&98.6&2&4 \\%&80\\
\\
30.88 & 0.27&0.088 &8.4 & 79.1 & 1 & 5 \\%& 60 \\ 
30.97&   -0.40&   0.241&    23.1&99.3&3&3 \\%&50 arc\\
30.98 &   0.08& 0.167& 16.3& 81.7 & 3&5 \\%&100\\ 
31.09&   -0.28&   0.076&     7.3&91.4&4&4 \\%&50\\
31.11&    0.13&   0.067&     6.4&81.0&3&5 \\%&70\\
\\
31.18&    0.22&   0.080&     7.7&81.7&3&6 \\%&90\\
31.19&   -0.09&   0.107&    10.3&81.7&2&7 \\%&40\\ 
31.41&    0.08&   0.130&    12.5&96.6&2&6 \\%&40\\ 
31.68&    0.20&   0.069&     6.6&99.2&3&4 \\%&80 elon\\
31.70&    0.05&   0.130&    12.5&96.0&3&11 \\%&40 arc\\ 
\\
31.79&   -0.38&   0.170&    16.3&98.0&2&2 \\%&70\\ 
31.92&    0.26&   0.078&     7.5&96.6&2&3 \\%&80\\

    \hline
\end{tabular}  
\label{tab-bubbles}
\end{center}
\end{table*}

\end{appendix}
\end{document}